\documentclass[11pt]{article}
\usepackage{graphicx}
\usepackage{amsfonts}
\usepackage{latexsym}
\usepackage{cite}

\begin{document}

\title{Aspects of M-5 brane world volume dynamics}
\author{David S. Berman\footnote{D.S.Berman@DAMTP.cam.ac.uk} \\ Department of Applied Mathematics 
and Theoretical Physics, \\Cambridge University, 
Wilberforce Road,\\ CB3 0WA Cambridge, England}

\maketitle
\begin{abstract}
This paper studies various aspects of the world volume dynamics of the M-theory five-brane, 
including: non-BPS solutions and solution generating symmetries; the scattering properties of world 
volume solutions; and the equivalence with probe brane dynamics.

\end{abstract}

\def    \bea    {\begin{eqnarray}}
\def    \eea    {\end{eqnarray}}
\def	\beq	{\begin{equation}} 
\def	\eeq	{\end{equation}}
\def	\lf	{\left (} 
\def	\rt	{\right )}
\def	\a	{\alpha} 
\def	\lm	{\lambda}
\def	\D	{\Delta} 
\def	\r	{\rho}
\def	\th	{\theta} 
\def	\rg	{\sqrt{g}} 
\def	\Slash	{\, \! \! \! \!}  
\def	\comma	{\; , \; \;} 
\def 	\pl 	{\partial} 
\def 	\del 	{\nabla}

\section{Introduction}

The M-theory five-brane still remains one of the most mysterious objects in M-theory. Its role in
M-theory is similar to that of a D-brane in string theory in that it is a submanifold on which open
membranes may end just as a D-brane is a submanifold on which open strings may end
\cite{openbranes}. The world volume dynamics of a D-brane, for slowly varying fields, are governed
by a Dirac Born-Infeld theory. A similar non-linear theory also exists for the five-brane world 
volume
fields \cite{m5eom}. Importantly, for D-branes there is also a description derived from open 
strings of the low energy 
dynamics (that is energies
below the string scale) of N coincident branes using U(N) Yang-Mills theory. 
This description of
D-branes has provided us with a rich insight in gauge theories. Unfortunately, there is no such
equivalent description for coincident M-theory branes. There are signs that a
description of the coincident five-brane theory will not be so simple. It is known from 
scattering calculations and black hole
thermodynamics that the number of low energy degrees of freedom of N coincident five-branes grows as
$N^3$ \cite{kleb,kostas}. There has been little insight so far as to reason for this growth in the 
number of degrees of freedom. For D-branes, where is growth is $N^2$, the answer is due to the open 
strings stretching 
between 
different  
D-branes becoming light as the D-branes become coincident which from the Yang-Mills point of 
view is simply the W-Bosons becoming massless so the gauge symmetry becomes enhanced 
to a U(N) theory. 

There is some hope that perhaps with a better understanding of the open membranes that stretch
between different five branes one might hope to understand the theory of coincident five-branes
using a similar construction. In this paper we will be less ambitious and attempt to study some 
basic aspects of how a open membranes end on five-branes using the world volume and brane probe
approaches. This approach was initiated for the BIon \cite{BIon} by \cite{callan} and then
numerous other works \cite{scat,lee,neil,kastor} investigated further scattering properties. This paper will 
concentrate on aspects relevant to the M-theory five brane. 

In particular, the first section of the paper will deal with symmetries of the five brane 
equations of motion that we will then use to generate a family of static non-BPS solutions. The 
second section of the paper will deal with scattering properties of world volume solutions and 
the equivalence with brane probes. This will allow us to speculate on the number of degrees of 
freedom carried by N self-dual strings in a theory with $N^{\prime}$ coincident five branes.

We should note that similar issues have been investigated recently in a series of papers by 
\cite{mans} using a linearised 
version of the field equations for the self dual string coupled to a two form and background 
scalar.

\section{Non-BPS solutions and solution generating symmetries}

The bosonic fields on the five brane include 5 scalar fields, $\phi^I$, $I=1..5$ and a two form 
potential $b$ whose field strength $H=db$ obeys a nonlinear self duality constraint. This self 
duality constraint together with the Bianchi identities imply the full equations of motion. 
There 
are various formulations for the five brane that impose this nonlinear 
constraint \cite{dima,per}. The form that we will 
find useful here is the SO(5) invariant formulation where one may write the self 
duality constraint using the following 
``Hamiltonian 
density" \cite{gary2}; (in fact strictly speaking what follows is is the Legendre transform of 
the 
Hamiltonian 
density also derived in \cite{gary2} and is expressed in terms of ${\it{electric}}$ fields 
only\footnote{We have made a trivial field redefinition $4H \rightarrow  H$ as compared with the 
formulation given in \cite{gary2}} )
\beq
{\cal{H}}=  \sqrt{\det (g_{ij} + i  E_{ij} )} -1 \, \, .
\eeq

Note the determinant is even in $E_{ij}$ so there is no imaginary part in the determinant.
The metric $g_{ij}$ is the induced metric given by the pull back of the background metric to 
the five-brane world volume. 
\beq
g_{ij} = \pl_i X^I \pl_j X^J G_{IJ}
\eeq

In so called static or Monge gauge where one identifies the world volume coordinates with the 
spacetime coordinates the induced metric becomes:
\beq
g_{ij}= G_{ij} + \pl_i \phi^A \pl_j \phi^B G_{AB}
\eeq
with the fields $\phi^A$ are the scalar fields on the five brane.

The so called {\it{electric}} and {\it{magnetic}} fields of the 
three form field strength $H=db$ are defined  respectively by:
\beq
 E_{ij} = H_{0ij} \quad  ; \quad B^{ij} =-{1 \over 6} \epsilon^{ijklm} H_{klm} \, \, .
\eeq

The nonlinear self duality constraint may now be expressed as a relation between the magnetic 
fields 
and the electric field given by:
\beq
B_{ij}=  { {\pl {\cal{H}} } \over {\pl E_{ij}} } = { { ( 1+{1 \over 2} TrE^2)E_{ij} + 
 (E^3)_{ij} } \over{ 
\sqrt{1+{1 \over 2} TrE^2 + U_i^2}}} \qquad \label{sd}
\eeq
where 
\beq
U_i = {1 \over 8} \epsilon_{ijrst} E^{jr} E^{st} \; 
\eeq
and we have implicitly included the scalar fields by using the induced metric in the 
contractions of the indices in the above equation.

One must also impose the Bianchi identities which in this formulation are equivalent to:
\beq
\pl_i B_{ij} =0 \, , \;  \;  \pl_t B_{ij} + {1 \over 2} \epsilon_{ijkrs} \pl_k E_{rs} =0  \; \; 
.
\eeq

These Bianchi identities combined with the self-duality constraint (\ref{sd}) imply the
equations of motion for the self dual field. This formulation of the five brane is discussed in
\cite{gary2,malc} and also connected to the various other equivalent formulations.

The equations of motion of the scalar fields are given by \cite{m5eom,per}:
\beq
G^{\mu\nu}\nabla^{(g)}_{\mu} \partial_{\nu} \phi^i =0
\label{sde}
\eeq
\begin{eqnarray}
\label{k}
G_{\mu\nu} &=& {1+K\over 2K}\left(g_{\mu\nu}+{\ell_p^6\over
4} {H}^2_{\mu\nu}\right)\ \, , \\
 K &=&\sqrt{1+{\ell_p^6 \over 24}{{H}}^2}\, ,\\
\end{eqnarray}
where $\nabla$ denotes a covariant derivative with respect to the induced metric $g_{\mu 
\nu}$. $G_{\mu \nu}$ is conformally related to the so called open membrane metric described in 
\cite{ommetric}.

We are interested in solving these equations to find the M-theory version of the non
supersymmetric BIon. We will make a string-like ansatz for the fields motivated by the
following physical arguments. On a D-brane where one has a one form potential there are point
like solutions. A five brane has a two form and so it is natural to look for string like
solutions. From the string/M-theory point of view this is just the usual addition of a new
dimension as one goes to strong coupling in string theory and strings become membranes.

We will look for solutions with a single excited scalar field $\phi$ and a non trivial two form
potential $b_{\mu \nu}$ in a flat background with metric $G_{IJ} =\eta_{IJ}$ as follows:
\beq 
\phi^1=\phi(r) \, , \quad   b_{01}=\chi(r) \, ,\quad b_{\phi \psi} = {q \over r^2}(\pm 1-cos 
\theta) \, .
\label{ansatz}
\eeq 
We have decomposed the six dimensional world volume in a ``2+4" split, that is $t$ and
$x_1$ will be the coordinates of the string world sheet and the remaining four coordinates
$(r,\theta,\phi,\psi)$ are the spherical coordinates of the space transverse to the string. This 
ansatz also of course makes a gauge choice for the form of the two form potential $b_{\mu \nu}$. 
In this gauge, the field
$\chi(r)$ can be thought of as an electric potential since the {\it{electric}} components of 
the field strength will be given by derivatives of $\chi$.
It is necessary to also include {\it{magnetic}} components of $b_{\mu \nu}$   so that the 
self-duality 
constraint will be obeyed. In fact, the ansatz made here includes a Dirac ``monopole" type 
potential for $b_{\phi \psi}$ to ensure the string will be magnetically charged.
With this ansatz, the Hamiltonian density greatly simplifies to:
\beq
{\cal{H}}= \sqrt{ 1+(\pl_r \phi(r))^2 -  (\pl_r \chi(r))^2 } - 1  \, \, .
\eeq

The first thing to notice is that the Hamiltonian density in this form has a global SO(1,1)  
symmetry rotating the $\phi$ and $\chi$ fields. This symmetry was pointed out for BIons in
\cite{BIon,gary3} and is similar to the Harrison transformations that relate a charged black hole to
a neutral black hole via a boost like transformation. In the case described here the
consequences of this symmetry are, given a solution with $(0,\chi)$ one can generate another
solution with $(\phi',\chi')$ given by:

\beq
\phi' = { v \over {\sqrt{1-v^2}} } \chi \, \; ,\quad \chi'= {1 \over {\sqrt{1-v^2} } } \chi \, \, . 
\label{trans}
\eeq

One should also remark that although we took a spherical ansatz (relevant to the case at 
hand) in fact this symmetry persists for any static configuration in static gauge. (This may be 
demonstrated using standard determinant manipulations).

In \cite{malc} a solution of the five brane field equations was found with $\phi=0$. This 
solution is given by:

\beq
E_r = \pl_r \chi(r) = { q \over { \sqrt{q^2 + r^6} } } \; . 
\eeq
Note that q is the same parameter that appears in the magnetic potential and is of course 
related the charge of string as can be seen by looking at the asymptotics of $E(r)$,
\beq 
 E(r) \rightarrow {q \over r^3} \, , \quad r \rightarrow \infty \, .
\eeq
Integrating this electric field strength over the $S^3$ at infinity will give an 
electric charge q. 
Note that the maximum field strength is as $r 
\rightarrow 0$, $E(r) 
\rightarrow 1$. This is the presence of a critical field strength that is typical of 
Born-Infeld theories (and indeed was part of the original motivation).

Substituting this field strength into the self duality condition (\ref{sd}) implies the magnetic 
field must be:
\beq
B_{1r} = {q \over r^3}
\eeq
which is consistent with the ansatz (\ref{ansatz}) given above; and implies that the solution 
has equal electric and magnetic charge, q. This charge q will be quantised using a 
generalisation of the usual Dirac argument for monopole charge quantisation. 

We can now use the transformation (\ref{trans}) to generate a continuous family of solutions.
This is illustrated in the above figure. The scalar field strength is drawn indicating how 
the brane becomes more and more deformed as v is increased and the electric field is ``boosted'' into the scalar field. These solutions will of course all be 
``spacelike " from the SO(1,1) point of view. One can
also find ``timelike" solutions that are disconnected from the these solutions where the
scalar field dominates. These are the five brane versions of the catenoid given in
\cite{gary3}. The ``lightlike" solutions ie. solutions that are fixed points under this SO(1,1)
symmetry are in fact the self-dual string solutions found by Lambert and West,
\cite{sdstring}. This solution (with charge N) is given by:

\beq \phi=  f = {N \over r^2} \, , \quad  {{H}}_{01p}= \partial_p f \, \quad {{H}}_{mnp} = \epsilon_{mnpq}\partial_q f \, \, . \label{sds} \eeq

 This string solution may be connected to the spacelike solutions by considering the 
divergent limit 
where the boost parameter, v is sent to one causing the scalar field strength to diverge.
This self dual string solution is one half BPS state from the
world volume theory point of view and corresponds to the emergence of the membrane from the 
five brane and as such has been the subject of much study.

Note,  the other solutions (that transform under this SO(1,1) symmetry)
are non-BPS. It is interesting to remark that it is the BPS solutions that are fixed points-
one may wonder whether this is a generic property of hidden solution generating symmetries in
M-theory \cite{west}.

It is not clear how to interpret these non-BPS solutions and they will certainly receive
quantum corrections since no supersymmetry is preserved. However, these solutions may be tuned
(for v near 1) to be very close to the BPS bound and as such it may prove interesting to study
in this near BPS limit.

\begin{figure}[]
\begin{center}
\includegraphics[width=4.1truecm]{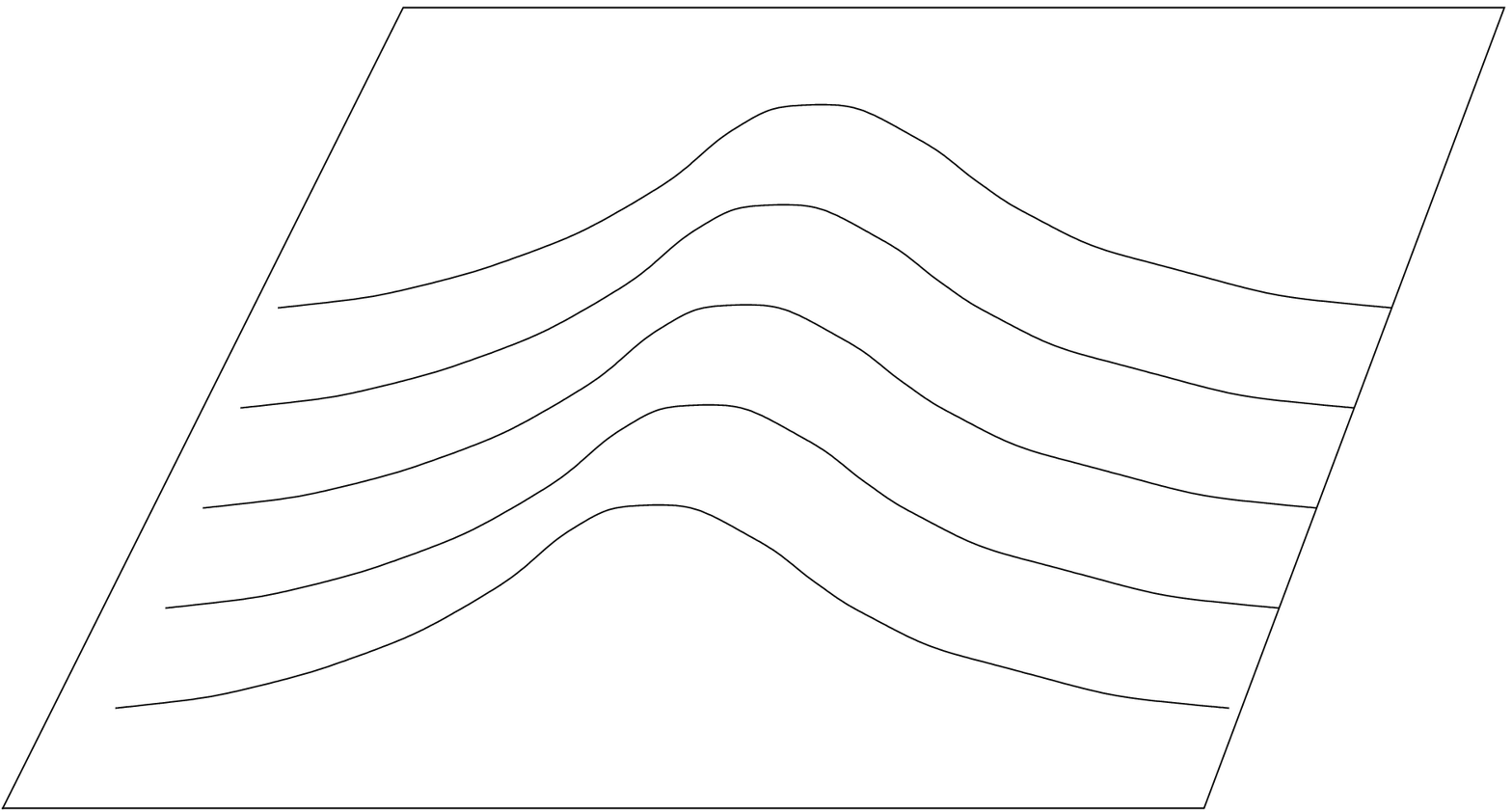}
\includegraphics[width=4.1truecm]{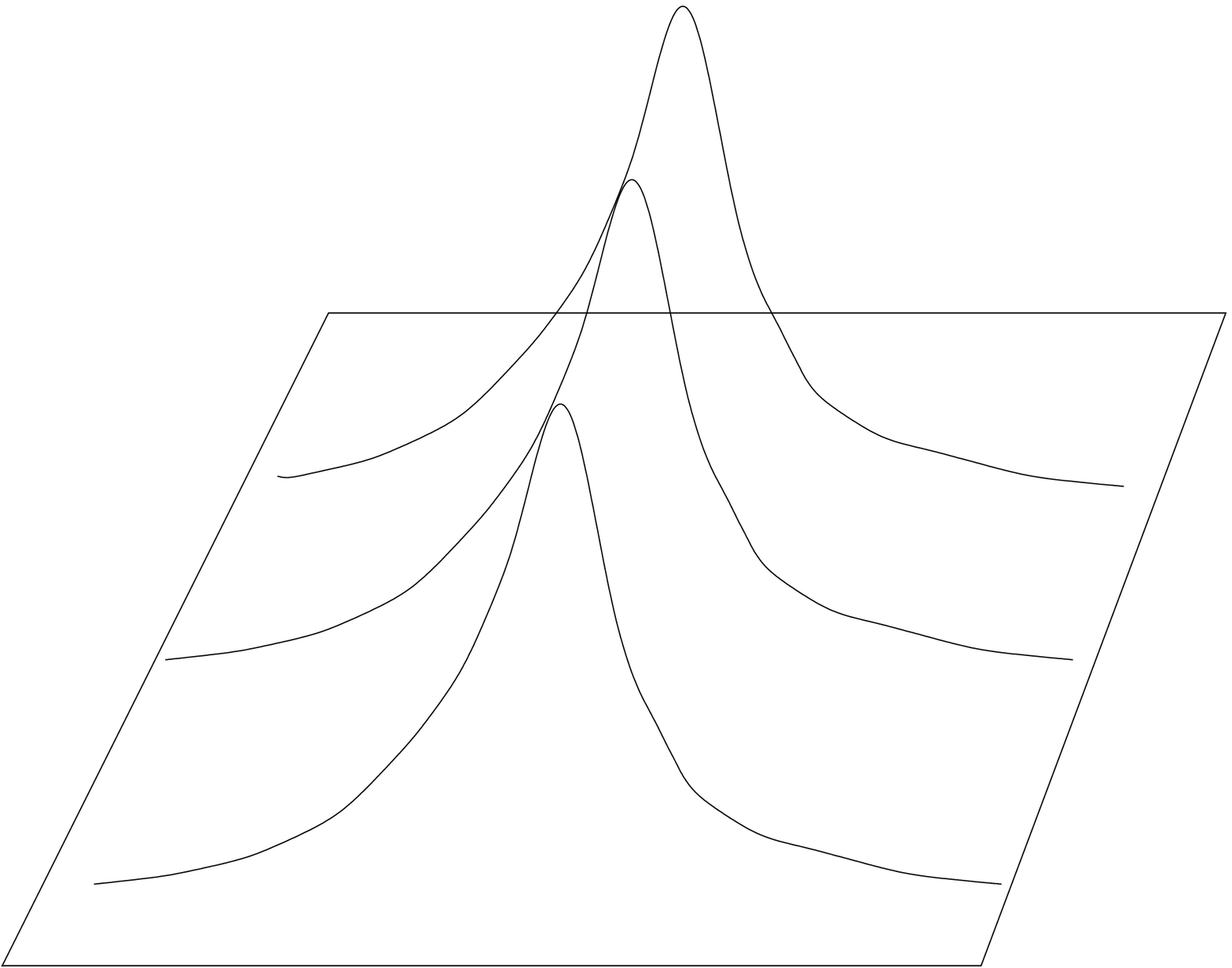}
\includegraphics[width=4.1truecm]{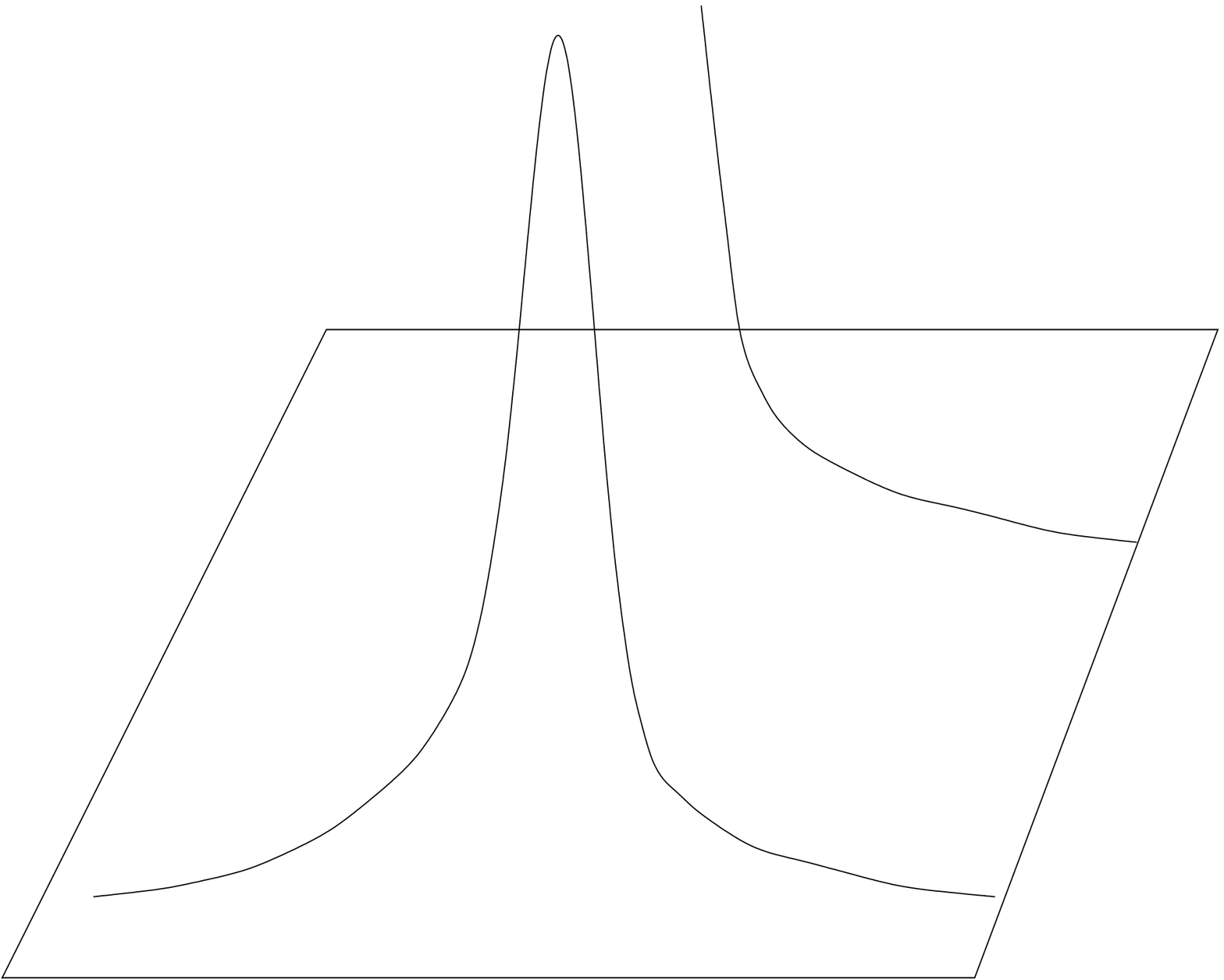}
\caption[]{Scalar field strength from small to large v}
\label{figure}
\end{center}
\end{figure}

\section{World Volume Scattering and Probe Brane Scattering}

In this section we will examine two related situations from which we will be able to study 
the scattering properties of membranes and five branes. The relationship between brane probes 
of supergravity solutions and world volume solitons was pointed out in \cite{paul}.  

The first is the situation we have already considered above, a single M-five brane with N
membranes ending on it. This may be described by a charge N, BPS self dual string solution 
living on the world volume of a single five brane.

The second is where we have $N^\prime$ five branes and a single membrane ending on it. This 
may 
be
described by a membrane probing a supergravity solution with $N^\prime$ units of five brane 
flux. (The 
solution will be given below).

These two scattering problems turn out to be mathematically equivalent once a suitable 
identification of parameters has been made. (This was pointed out for the case of BPS BIons on 
D-branes in \cite{scat}.)

Let us first write down the equation of motion an s-wave scalar fluctuation of energy $\omega$
that is transverse to both the M-5 brane and the self dual string. This is obtained by expanding
the scalar field equations (\ref{sde}) to linear order around the charge N self dual string solution (\ref{sds}).
This was first described in \cite{me},
\beq
\left( \rho^{-3} { d \over {d\rho}} \rho^3 {d \over {d \rho}} +1 + {{R^6 \omega^6} \over 
{\rho^6}} 
\right) \phi(\rho) =0
\eeq
where $\rho=r \omega$ the dimensionless distance transverse to the string and 
\beq
R^3 = N l_p^3  \quad .
\eeq

\begin{figure}[]
\begin{center}
\includegraphics[width=7.truecm]{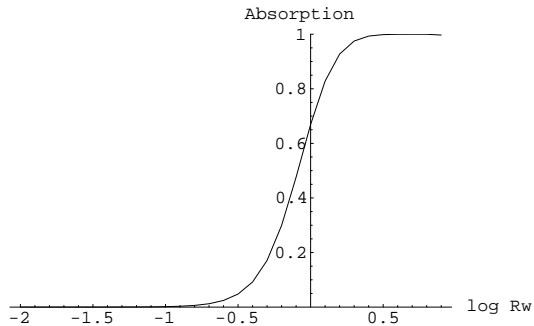}
\caption[]{Absoption as a function of log R $\omega$}
\label{figure}
\end{center}
\end{figure}

To calculate the absorption cross section of the string one calculates solutions for the interior and exterior regions (exactly what we we mean by interior and exterior is defined below) and then match the solutions. One may then take the ratio of the incoming fluxes in the interior region to the exterior region to obtain a (dimensionless) absorption cross section. 

The asymptotic solution (for the outer region where $\rho >> R \omega$) are given by 
simple Bessel functions.

One may then change to variables that are suited to the inner region of the solution 
(essentially one changes coordinates to those adapted to the membrane emerging from the five 
brane).

\beq
z= {{ (R \omega)^3 } \over {2 \rho^2 }}
\eeq

In these variables the equation of motion becomes:
\beq
\left(  {{d^2 } \over {dz^2 }} + {{({1 \over 2} R \omega)^3} \over {z^3}} +1 \right) \phi(z) 
=0
\eeq

Solutions in the interior region  (where $z>>R \omega$) corresponding to the membrane 
fluctuations maybe written in terms of simple trigometric functions. 

As shown in \cite{me}, the solutions in these two asymptotic regions can be matched provided $R 
\omega <<1$ to obtain a low energy scattering (dimensionless) cross section given by:
\beq
\sigma \sim  (R \omega)^3 \sim N 
\eeq

The implication of this calculation is that the number of degrees of freedom of N coincident 
self dual strings is proportional to 
N. (Recall how for a D-brane the cross section scattering of bulk supergravity fields is 
proportional to $N^2$ which of 
course gives the correct number of degrees of freedom for a U(N) gauge theory 
\cite{kleb}. It is slightly 
puzzling when one compares this to the absorption cross section of a membrane in eleven 
dimensional supergravity where the absorption cross section of Q membranes is proportional to 
$Q^{3 \over 2}$.

To calculate the absorption cross section for generic $R \omega$ one may also solve these
equations numerically. This was done using a simple Mathematica script. For small $R\omega$ the numerical solutions matched the analytic solutions of the asymptotic regions as they must. As $R \omega$ approaches one the absorption greatly increases till for high energies the waves are completely absorbed.
This is shown in figure 2, where
we have plotted the dimensionless absorption cross section against $\rm{log}(R\omega)$. As $R\omega$ nears 1 the membrane becomes fully absorbing, numerically confirming the analysis given in
\cite{callan} for BIons.

We will now examine the alternative situation of one membrane ending
on $N^\prime$  five branes by placing a probe membrane in a
supergravity background describing $N^\prime$  coincident five
branes. This was first  discussed in \cite{paul} a similar scattering
calculation was done in the string/D-brane  context by \cite{lee,kastor}.

The five brane background is given by:  \bea ds^2 &=& H^{- {1\over3}}
(-dx_0^2+...+dx_5^2) + H^{{2\over3}} (dx_6^2+...+dx_{10}^2) \\
F_{mnpq}&=& \epsilon_{mnpqr} \pl_r H  \eea where  \beq H=1 +
{{{\tilde{R}}^3} \over {r^3}} \, \, , \qquad  \, {\tilde{R}}^3=
N^{\prime} l_p^3 \, \, , \qquad    r^2=x_6^2+..+x_{10}^2 \eeq The
membrane probe action is \beq S=\int d^3 \sigma
\sqrt{-{\rm{det}}(G_{mn})} + \int (f^*C)_{mnp} \epsilon ^{mnp}
\label{ng} \eeq Where $f^*C$ denotes the pull back of the 3-form
potential C and $G_{mn}$ is the induced  metric given by the pull back
from the space-time metric.

We will consider a membrane stretched radially outward from the five
brane stack. This can be  described after picking suitable coordinates
by: \beq \sigma^0 = x^0 \comma \sigma^1=x^1 \comma \sigma^2=x^{10} \,
\, . \label{m2} \eeq It is trivial to check that this configuration
solves the equations of motion given by varying  the above action.

We wish to examine small fluctuations of this configuration. That is
we will do a background  field expansion around the above solution as
follows, \beq X^{\alpha}(\sigma)= X_{cl}^{\alpha} +x^{\alpha}(\sigma)
\label{bf} \eeq where $X_{cl}^{\alpha}$ denote the classical
configuration described by equation (\ref{m2}) above and  $x^{\alpha}(\sigma) $
is the fluctuation field.

We then decompose the spacetime coordinates into  those longitudinal
to the  five brane, transverse to the five brane and membrane and
finally transverse to the membrane  but longitudinal to the membrane
as follows, \beq x^{\alpha} \rightarrow (x^a,x^{\mu},x^{10})
\label{decomp} \eeq where $a=0..5,\mu=6..9$.  We then derive the
equations of motion for the s-wave fluctuations transverse to both the
five  brane and membrane by inserting the background field expansion
(\ref{bf}) with background  field  (\ref{m2}) and coordinate
decomposition (\ref{decomp}) into the action (\ref{ng}). Then,
examining  the fluctuations transverse to both branes implies the
equation of motion for  $x^{\mu}=x^{\mu}(\sigma^2,t)$ is given by:
\beq \pl_r^2 x^{\mu} - H(r) \pl_t^2 x^{\mu} =0 \, \, .  \eeq Solutions
with energy $\omega$, are given by $x^{\mu}=e^{-i\omega t} x^{\mu}
(z)$ where we have introduced the dimensionless distance $z=r \omega$,
with $x^{\mu}$ now obeying the following  equation of motion: \beq
\left( {{d^2} \over {dz^2}} + 1 + {{({\tilde{R}} \omega)^3} \over
{z^3}} \right) x^{\mu} (z) =0  \eeq  Now we remark that this is the
identical equation to that derived using the background field
expansion of the five brane equations of motion around the self dual
string solution ie. the previous set  up described earlier in the
section. This should not come as a surprise since the scattering
processes we are considering are clearly related. Using the previous
results for solving this equation implies  the dimensionless
absorption crossection in this case is: \beq \sigma \sim {\tilde{R}}^3
\sim N^{\prime} \, \, , \eeq where $N^{\prime}$ is the number of five
branes. This means that in this process the absorption cross  section
and hence the number of degrees of freedom available for the probe
membrane to scatter into is  proportional  to the number of  five
branes.

 This leads us to conjecture that for a generic configuration of
$N^{\prime}$  five branes and N membranes the cross section would be
$N \times N^{\prime}$. Note, this is  different to the case described
in \cite{lee} with D-branes and strings where the identification
between parameters is more complicated. If one wishes to  interpret
the scattering as being related to the number of degrees of freedom
accessible to  the object then this would imply the number of low energy degrees
of freedom of N self dual strings  contained in $N^{\prime}$ five
branes goes as $N \times N^{\prime}$.

The scattering of the non-BPS solutions found earlier maybe calculated
using a simple generalisation of the above and shows that the
scattering cross section decreases for the non-BPS solution as
compared with the BPS solution in concordance with the results for
BIons discussed in \cite{nonbpsscat} and of course the N dependence
remains the same.

\section{Conclusions}

This paper has attempted to study various aspects of how membranes end
on five-branes. These investigations were originally prompted by the
lack of a current understanding of a ``non-abelian'' five-brane theory
and also by previous work concerning the near horizon of the self-dual
string \cite{me}. In \cite{me} there was a conjectured low energy
duality between a five-brane with $AdS_3 \times S^3$ geometry and the
theory of N coincident self-dual strings. From the scattering
calculations described above we have seen how the number of low energy degrees of freedom
of the self-dual string depend on N (the self-dual string charge) and
$N^{\prime}$ (the number of five branes); the answer may not be very
surprising but given the previous surprises in M-theory (eg. the
$Q^{3/2}$ dependence of the membrane cross section) it is worth
obtaining the answer. This result also seems to be consistent with the
intuition gained by looking at $N^\prime$ five branes intersecting
with N five branes on a string (even though that set up is quite
different to the one described here one might imagine a relationship) for
that situation, studied in \cite{neil2} anomaly arguments also imply
$N^\prime \times N$ degrees of freedom for the string intersection.

It would also be useful to make a quantitative comparison with the
model of self-dual strings described in \cite{mans} where Thompson
scattering was calculated using an alternative description of the
five-brane coupled to a self-dual string. 
 
 We have also constructed non-BPS solutions of the five-brane. It is
 not clear what role they play in M-theory or indeed what the non-BPS
 BIons play in string theory.  Since these objects will have large
 quantum corrections one perhaps shouldn't give them too much credence
 however (using the SO(1,1) symmetry) one may construct near BPS
 solutions where one can take the solution to be as near BPS as
 required. These near BPS solutions may well prove interesting to
 study just as near BPS solutions have proved more tractable in
 theories with gravity. One possible interpretation of these non-BPS
 solutions \footnote{We are grateful to J Maldacena for providing this interpretation.} is that they describe a situation with $n$ membranes ending
 on one side of the five brane and $m$ membranes ending on the other
 side. The scalar charge is then related to $m-n$ but the two form
 charge is related to $m+n$. The BPS condition which equates the
 scalar and two form charges then implies that $n=0$ and one recovers
 the usual charge $m$ self dual string solution.

\section{Acknowledgments}

It is a pleasure to acknowledge the numerous useful discussions I've
benefited from with Robert Helling, Malcom Perry, Juan Maldacena,
Dmitri Sorokin, Paul Townsend, Toby Wiseman and especially Gary
Gibbons.  The author is supported by an EPSRC advanced fellowship,
GR/R75373/01, project ``Aspects of M-theory Interactions''.

\end{document}